\documentstyle[preprint,aps]{revtex}
\input epsf.tex
\def\DESepsf(#1 width #2){\epsfxsize=#2 \epsfbox{#1}}
\begin{document}
\preprint{\vbox{\hbox{}}}
\draft
\title{Non-Spectator Contributions to Inclusive  Charmless $B$ Decays}
\author{Wu-Sheng Dai$^{1,2}$, Xiao-Gang He$^{3}$, Xue-Qian Li$^{1,2}$
and Gang Zhao$^{1,2}$}

\address{1. CCAST (World Laboratory), P.O. Box 8730, Beijing 10080\\
2. Department of Physics, Nankai University, Tianjin, 300071\\
3. Department of Physics, National Taiwan University, Taipei, 10764
}

\date{November 1998}
\maketitle
\begin{abstract}
The light quarks inside $B$
 mesons are usually treated as spectators and 
do not affect the decay rates which are assumed to be purely 
due to b quark decays.
In this paper we calculate the non-spectator 
contributions to inclusive 
charmless $B$ decays due to the spectator effects. 
We find that the non-spectator contributions to 
the branching ratio for $\bar B^0$ are small ($<2\times 10^{-4}$),
but the contributions to 
$\Delta S = 0$ and  
$\Delta S = -1$, $B^-$ decay branching ratios can be as large as 
$-7.5\times 10^{-4}$ and $2\times 10^{-3}$, respectively.
These contributions 
may play 
an important role in rare charmless $B$ decays.
\end{abstract}

Studies of B-physics greatly enrich our
understanding of 
the interactions involving heavy quarks.
In the heavy quark limit,
the light quarks inside the B mesons are treated as 
spectators which do not affect the decay rates\cite{0}. 
If the heavy quark is very heavy compared to the $QCD$ scale $\Lambda_{QCD}$, 
this approximation is a good one because 
the effect from the light quark
is suppressed by two powers in heavy quark mass compared with that of the
three body decays of b quark. However, in reality the b quark is not infinitively
heavy, the suppression factor proportional to $\Lambda^2_{QCD}/m_b^2$ may be 
overcomed by the enhancement factor of
 $16\pi^2$ in phase space because the spectator effects induced
decays
are two body decays\cite{1}. 
Therefore spectators may affect in some way the branching ratios,
especially in rare $B$ decays.
We will refer the effects due to the light spectator quark 
inside the $B$ meson as non-spectator effects.

It has been shown that the dominant non-spectator  
effects at tree level can play
an important role in  the missing charm and the $\Lambda_b$ lifetime
problems\cite{1,2}.
The non-spectator effects also have important implications for
 exclusive decays\cite{3} where the corresponding effects are usually 
called the annihilation effects. 
These effects are usually assumed to be
small and are neglected. It has been shown 
that if the annihilation contributions to $B^-\to \bar K^0 \pi^-$ 
are really small,
it will  be possible 
to determine one of the fundamental parameter $\gamma$ in the
unitarity triangle by measuring several $B$ decay modes\cite{4}. 
It is
however very difficult to calculate the annihilation contributions for 
exclusive decays.
Without a reliable calculation, we have to find some ways to experimentally 
test if 
the annihilation contributions are small\cite{3}. 
On the other hand the analogous contribution, the non-spectator 
contribution, in inclusive B decays may be easier to study.
From this study one may also obtain some useful information 
about annihilation contributions for exclusive decays.
In this paper we will carry out a calculation for the non-spectator 
contributions to the inclusive charmless $B$ meson decays in the 
Standard Model.

In the Standard Model the quark level effective Hamiltonian responsible for
charmless $B$ decays are given by\cite{5}

\begin{eqnarray}
H_{eff} = {4G_F\over \sqrt{2}} [V_{ub}V_{f_2q}^*(c_1O_1(q)
+c_2 O_2(q))
-V_{tb}V_{tq}^*\sum_{i=3-6} c_i O_i(q)],
\end{eqnarray}
where
\begin{eqnarray}
&&O_{1}(q)=\bar u \gamma_\mu L u \bar q \gamma^\mu L b, \;\; 
O_2(q) = \bar q \gamma_\mu L u \bar u \gamma^\mu 
L  b,\nonumber\\
&&O_{3,5}(q) = \bar q \gamma_\mu L b \bar q' \gamma^\mu L(R) q',
\;\;
O_{4,6}(q)= \bar q_{\alpha} \gamma_\mu L b_{\beta}
\bar q'_\beta \gamma^\mu L(R) q'_\alpha,
\end{eqnarray}
where $L(R) = (1\mp \gamma_5)/2$,
 $q'$ is summed over $u,d,s$, and $q$ can be $d$ or $s$ depending on whether 
the processes are $\Delta S = 0$ or $\Delta S = -1$.
In the above we have neglected electroweak penguin contributions.
In our later discussions we will use the Wilson coefficients evaluated in 
Ref.\cite{6}.

There are several quark level processes correspond to 
non-spectator contributions.  
They are shown in Figs. 1 and 2. Figs. 1 and 2 are induced by tree
and penguin operators, respectively. 
We will refer the contributions from Figs. 1a, 1b, 2a, and 2b as 
Usual Non-spectator (UN) contributions, while refer those from 
Figs. 1c, 2c and 2d, which  are due to Pauli
interferences, as Pauli Interference (PI) contributions. The PI 
contributions reduce the branching ratios.
Using the optical theorem, the inclusive decay width of $B$ can be written as 
the forward matrix element of the imaginary part of the transition operator
$T$,

\begin{eqnarray}
\Gamma(B) = {1\over m_B}Im \langle B|T|B\rangle 
= {1\over 2m_B}\langle B|\tilde \Gamma|B\rangle,
\end{eqnarray}
where $T$ is given by
\begin{eqnarray}
T = i \int d^4x T\{H_{eff}(x), H_{eff}(0)\}.
\end{eqnarray}

Evaluating diagrams in Figs. 1 and 2 and neglecting 
terms proportional to light quark masses, we obtain

\begin{eqnarray}
\tilde \Gamma^q(1a) &=&
-{2\over 3}{G_F^2|V_{ub}V_{uq}^*|^2m_b^2\over \pi}[(N({c_1\over N}+c_2)^2 
(O^u_{V-A}-O^u_{S-P})
+2c_1^2(T^u_{V-A}-T^u_{S-P})],\nonumber\\
\tilde \Gamma(1b) &=&
-{2\over 3}{G_F^2|V_{ub}V_{ud}^*|^2m_b^2\over \pi}[(N(c_1+ {c_2\over N})^2 
(O^d_{V-A}-O^d_{S-P})
+2c_2^2(T^d_{V-A}-T^d_{S-P})],\nonumber\\
\tilde \Gamma^q(1c)	
&=&2{G_F^2|V_{ub}V_{uq}^*|^2m_b^2 
\over \pi}[(2c_1c_2 + {1\over N}(c_1^2+c_2^2)) 
O^{u}_{V-A} +2 (c_1^2+c_2^2) T^{u}_{V-A}], \nonumber\\
\tilde \Gamma(2a) &=& -{2\over 3}{G_F^2|V_{tb}V_{td}^*|^2m_b^2\over \pi} 
 3[N((c_3+{c_4\over N})^2+(c_5+{c_6\over N})^2)
 (O^d_{V-A}-O^d_{S-P})\nonumber\\
&+& 
2(c_4^2+c_6^2)(T^d_{V-P}
-T^{d}_{S-P})],\nonumber\\
\tilde \Gamma^q(2b) &=& -{2\over 3}{G_F^2|V_{tb}V_{tq}^*|^2m_b^2 \over \pi} 
 [N({c_3\over N}+c_4)^2
 (O^u_{V-A}-O^u_{S-P}) +
2c_3^2(T^u_{V-P}
-T^{u}_{S-P})\nonumber\\
&-&6N({c_5\over N}+c_6)^2\tilde O^{u}_{S-P} - 12 c_5^2
\tilde T^{u}_{S-P}],\nonumber\\
\tilde \Gamma^q(2c)
&=&2{G_F^2|V_{tb}V_{tq}^*|^2 m_b^2\over \pi}[(2c_3c_4 + {1\over N}(c_3^2
+ c_4^2)) O^{u}_{V-A} +2 (c_3^2+c_4^2)T^{u}_{V-A}
\nonumber\\
&-&(2c_5c_6 + {1\over N}(c_5^2+c_6^2))
({2\over 3}\tilde O^{u}_{S-P}-{1\over 6} \tilde O^{u}_{V-A})-
 2 (c_5^2+c_6^2)({2\over 3}
 \tilde T^{u}_{S-P}-{1\over 6} \tilde T^{u}_{V-A})],\nonumber\\
\tilde \Gamma^q(2d)
&=&2{G_F^2|V_{tb}V_{tq}^*|^2m_b^2 \over \pi}[(2c_3c_4 + {1\over N}(c_3^2
+ c_4^2)) O^{d}_{V-A} +2 (c_3^2+c_4^2))T^{d}_{V-A}
\nonumber\\
&-&(2c_5c_6 + {1\over N}(c_5^2+c_6^2))
({2\over 3}\tilde O^{d}_{S-P}-{1\over 6} \tilde O^{d}_{V-A})-
 2 (c_5^2+c_6^2)({2\over 3}
 \tilde T^{d}_{S-P}-{1\over 6} \tilde T^{d}_{V-A})].
\end{eqnarray}

\begin{eqnarray}
O^q_{V-A}& =& \bar b \gamma_\mu L q \bar q\gamma^\mu Lb,\;\;
O^q_{S-P}= \bar b  L q \bar q R b,\nonumber\\
T^q_{V-A}&=&\bar b \gamma_\mu LT^aq \bar q\gamma^\mu LT^ab,\;\;
T^q_{S-P}=\bar b  L T^a q \bar q RT^a b,\nonumber\\
\tilde O^q_{V-A}&=&\bar b \gamma_\mu R q \bar q\gamma^\mu Rb,\;\;
\tilde O^q_{S-P}= \bar b  R q \bar q L b,\nonumber\\
\tilde T^q_{V-A}&=&\bar b \gamma_\mu RT^a q \bar q\gamma^\mu RT^a b,\;\;
\tilde T^q_{S-P} = \bar b  R T^a q \bar q L T^a b.
\end{eqnarray}

From the above we can obtain the non-spectator contributions to 
$B^-$ and $\bar B^0$ decays using eq. (3). For $\Delta S = 0$ decays, we have

\begin{eqnarray}
\Gamma(B^-\to X)
&=&\Gamma_0 |V_{ub}V_{ud}^*|^2\{[N({c_1\over N}+c_2)^2 (B_2-B_1)
+2c_1^2(\varepsilon_2-\varepsilon_1)]\nonumber\\
&+&3[(2c_1c_2+{1\over N}(c_1^2+c_2^2))B_1 + 2(c_1^2+c_2^2)\varepsilon_1]\}
\nonumber\\
&+&\Gamma_0|V_{tb}V_{td}^*|^2\{ [N({c_3\over N}+c_4)^2 (B_2-B_1) +
2c_3^2(\varepsilon_2
-\varepsilon_1)\nonumber\\
&+&6N({c_5\over N}+c_6)^2\tilde B_2 + 12 c_5^2
\tilde \varepsilon_2]\nonumber\\
&+&3[(2c_3c_4+{1\over N}(c_3^2+c_4^2))B_1 + 2(c_3^2+c_4^2)\varepsilon_1
\nonumber\\
&-&(2c_5c_6 + {1\over N}(c_5^2+c_6^2))({2\over 3}\tilde B_2 - {1\over 6}
\tilde B_1)
- 2(c_5^2+c_6^2) ({2\over 3} \tilde \varepsilon_2 - {1\over 6}
\tilde \varepsilon_1)]\},\nonumber\\
\Gamma(\bar B^0\to X)
&=&\Gamma_0 |V_{ub}V_{ud}^*|^2[N(c_1+{c_2\over N})^2 (B_2-B_1)
+2c_2^2(\varepsilon_2-\varepsilon_1)]\nonumber\\
&+&3\Gamma_0|V_{tb}V_{td}^*|^2 [N((c_3+{c_4\over N})^2
+(c_5+{c_6\over N})^2) (B_2-B_1)\nonumber\\
 &+&
2(c_4^2+c_6^2)(\varepsilon_2
-\varepsilon_1)]\nonumber\\
&+&3[(2c_3c_4+{1\over N}(c_3^2+c_4^2))B_1 + 2(c_3^2+c_4^2)\varepsilon_1
\nonumber\\
&-&(2c_5c_6 + {1\over N}(c_5^2+c_6^2))({2\over 3}\tilde B_2 - {1\over 6}
\tilde B_1)
- 2(c_5^2+c_6^2) ({2\over 3} \tilde \varepsilon_2 - {1\over 6}
\tilde \varepsilon_1)]\}
\end{eqnarray}

And for $\Delta S = -1$ decays, we have
 
\begin{eqnarray}
\Gamma(B^-\to X_s)
&=&\Gamma_0 |V_{ub}V_{us}|^2\{[(N({c_1\over N}+c_2)^2 (B_2-B_1)
+2c_1^2(\varepsilon_2-\varepsilon_1)]\nonumber\\
&+&3[(2c_1c_2+{1\over N}(c_1^2+c_2^2))B_1 + 2(c_1^2+c_2^2)\varepsilon_1]\}
\nonumber\\
&+&\Gamma_0|V_{tb}V_{ts}^*|^2\{ [N({c_3\over N}+c_4)^2 (B_2-B_1) +
2c_3^2(\varepsilon_2
-\varepsilon_1)\nonumber\\
&+&6N({c_5\over N}+c_6)^2\tilde B_2 + 12 c_5^2
\tilde \varepsilon_2]\nonumber\\
&+&3[(2c_3c_4+{1\over N}(c_3^2+c_4^2))B_1 + 2(c_3^2+c_4^2)\varepsilon_1
\nonumber\\
&-&(2c_5c_6 + {1\over N}(c_5^2+c_6^2)({2\over 3}\tilde B_2 - {1\over 6}
\tilde B_1)
-2(c_5^2+c_6^2) ({2\over 3} \tilde \varepsilon_2 - {1\over 6}
\tilde \varepsilon_1)]\},\nonumber\\
\Gamma(\bar B^0\to X_s)&=&
3\Gamma_0|V_{tb}V_{ts}^*|^2
[(2c_3c_4+{1\over N}(c_3^2+c_4^2))B_1 + 2(c_3^2+c_4^2)\varepsilon_1
\nonumber\\
&-&(2c_5c_6 + {1\over N}(c_5^2+c_6^2)({2\over 3}\tilde B_2 - {1\over 6}
\tilde B_1)
-2(c_5^2+c_6^2) ({2\over 3} \tilde \varepsilon_2 - {1\over 6}
\tilde \varepsilon_1)],
\end{eqnarray}
where $\Gamma_0 =G_F^2 m_b^2m_Bf^2_B/12 \pi$, 
and the parameters $B_i(\tilde B_i)$ and
$\varepsilon_i (\tilde \varepsilon_i)$ are defined as follows,

\begin{eqnarray}
&&<B|\bar b \gamma_\mu L q \bar q\gamma^\mu Lb|B> = {f_B^2 m_B^2\over 4} B_1,\;\;
<B|\bar b  L q \bar q R b|B> = {f_B^2 m_B^2\over 4} B_2,\nonumber\\
&&<B|\bar b \gamma_\mu LT^aq \bar q\gamma^\mu LT^ab|B> = {f_B^2 m_B^2\over 4} 
\varepsilon_1,\;\;
<B|\bar b  L T^a q \bar q RT^a b|B> = {f_B^2 m_B^2\over 4} \varepsilon_2,\nonumber\\
&&<B|\bar b \gamma_\mu R q \bar q\gamma^\mu Rb|B> = {f_B^2 m_B^2\over 4} \tilde B_1,\;\
<B|\bar b  R q \bar q L b|B> ={f_B^2 m_B^2\over 4} \tilde B_2,\nonumber\\
&&<B|\bar b \gamma_\mu RT^a q \bar q\gamma^\mu RT^ab|B> = {f_B^2 m_B^2\over 4} \tilde 
\varepsilon_1,\;\;
<B|\bar b  R T^a q \bar q L T^a b|B> = {f_B^2 m_B^2\over 4} \tilde \varepsilon_2.
\end{eqnarray}
The above definitions are inspired by the factorization
approximation calculation. In this approximation, $B_i = \tilde B_i = 1$ and
$\varepsilon_i = \tilde 
\varepsilon_i = 0$. Conservation of strong interaction implies
$B_i=\tilde B_i$ and $\varepsilon_i =\tilde \varepsilon_i$.
There have been some attempts to calculate
$\varepsilon_i$ using the $QCD$
sum rules\cite{7}. The numbers obtained are
$\varepsilon_1 \approx -0.15$ and $\varepsilon_2 \approx 0$.
To see how the results change with hadronic parameters,
we will take two sets of representative
values in our 
latter analyses: a) $B_i(\tilde B_i) = 1$,
$\varepsilon_i(\tilde \varepsilon_i) = 0$, and
b) $B_i(\tilde B_i) = 1$, $\varepsilon_1(\tilde \varepsilon_1) = -0.15$, 
$\varepsilon_2 (\tilde{\varepsilon}_2) = 0$.
The numerical values also depend on the values of several KM matrix 
elements and the B decay constant $f_B$. We will use the values for these
parameters given in the table caption for illustration. One can easily find
out the changes using eq. (7) and (8) for other values of the parameters 
involved.
Our numerical results for the branching ratios are given in Table 1.

In the factorization approximation, $\bar B^0$ decays do not
receive tree non-spectator contributions and $B^-$ decays only receive
PI contributions. Note also that 
even non-factorizable contributions are included, the tree PI
 terms do not contribute to $\bar B^0$ decays.

 For $\Delta S =0$
decays, the non-spectator contributions are dominated by the PI 
contribution at tree level which reduces the total branching ratio
by $-2.8\times 10^{-4}$ in the factorization approximation. 
Non-factorizable effects can change the 
situation significantly. Using $\varepsilon_1 = -0.15$ and keeping the other 
parameters unchanged, the total branching 
ratio can be reduced by $-7.6\times 10^{-4}$ for $B^-$ decays, and the 
branching ratio for $\bar B^0$ decay can be increased by $1.4\times 10^{-4}$ 
from UN contributions. 
These effects
although small for total branching ratios, 
but when study rare charmless decays,
it may play an important role. For example for even just a few percent of the
non-spectator
 contributions find their way to $B^-\to \pi^- \pi^0$, the branching
ratio for this exclusive decay can change more than 10\%.
The penguin
non-spectator contributions are much smaller ($<2\times 10^{-5}$) and do not 
play an important role.

The branching ratios for charmless and $\Delta S =0$, $B^-$ and 
$\bar B^0$ decays are of order $5\times 10^{-3}$
 from the three body b quark decays\cite{8}. The non-spectator 
contributions can reach $6\% \sim 12\%$ 
of the main contributions. Also the non-spectator
contributions decrease the branching ratio for $B^-$ and increase the
branching ratio for $\bar B^0$. Experiments in the future may be able to 
observe these effects.

For $\Delta S =-1$ $B$ meson decays, the roles played by tree and penguin 
non-spectator contributions are reversed. The non-spectator contributions can
increase the branching ratio for $B^-$ decays by 
as much as $1.9\times 10^{-3}$.
$\bar B^0$ 
decays only receive PI  contributions and the
branching ratio can be reduced at a few times of $10^{-4}$ level.
The tree non-spectator contributions to $B^-$ can reduce the total branching 
ratio by a few times $10^{-5}$. This is small, but may still play an important
role in the study of rare $B^-\to \bar K^0 \pi^-$ decay. Usually the
tree amplitude for this decay is assumed to be extremely small. If this is
true this decay mode can be used in combination with several other
decays, for example, $B^-\to K^- \pi^0$ and $B^-\to \pi^-\pi^0$ decays,
 to determine the parameter $\gamma$ in the KM unitarity triangle\cite{4}. 
If 
annihilation contribution to the branching ratio of 
$B^-\to \bar K^0 \pi^-$ decay is 
large $O(10^{-6})$ (10\% of the inclusive non-spectator contributions), 
it will cause large uncertainty in the 
determination of the angle $\gamma$. 
However at present we do have reliable ways to
make a vigorous calculation. More detailed study is required.

The non-spectator contributions to the total branching ratios for
 $\Delta S = -1$ $B$ mesons are small, but can be 20\% of the three body
b quark decay contribution (1\%)\cite{8} to charmless $B$ meson decays. 
This contribution can not be neglected. Also the
non-spectator contributions affect $B^-$ and $\bar B^0$ differently. 
Experiments in the future may observed such effects.
Another important feature of the non-spectator contributions
is
that the decay modes induced by non-spectator effects are two body type. Their 
effect may be more eminent if kinematic cut requiring the decays to be two 
hard jets with small invariant masses is applied.

We would like to comment on the penguin 
non-spectator effects on  
the missing charm and the $\Lambda_b$ lifetime problems before conclude the 
paper. For this
discussion, we will also need to 
consider final states with charm quarks because this is the main non-spectator 
contribution. It has been
shown that the tree non-spectator contributions can be important\cite{1}.
From factorization calculation without the PI 
contributions, one might think that 
the penguin contributions to be important.
In this case, the dominant 
tree
non-spectator contribution is from $b\bar d \to c \bar u$ which is 
proportional to $(c_1+c_2/N)^2$, whereas the penguin non-spectator contribution
is dominated by $b\bar d \to s\bar d$ as can be seen from our previous 
discussions. Because the accidental cancellation between $c_1$ and $c_2/N$ for
$N=3$, the penguin contribution is about 20 times of the tree contribution in 
the factorization approximation. However, this result is very sensitive to
the values for $\varepsilon_i$. With $\varepsilon_1 = -0.15$ for example, 
the tree annihilation is 10 times larger than penguin contributions. 
Also, when PI contributions are included, the main tree 
contribution is proportional to
$c_1 c_2$ which is not small. The
tree non-spectator contributions turn out to be 15 to 30 times larger than
the penguin non-spectator contributions. 

We also carried out a detailed calculation for the penguin non-spectator 
contributions to the $\Lambda_b$ lifetime. We again find that the 
penguin non-spectator
 contributions are only a few percent of the tree non-spectator 
contributions. 
We conclude that the penguin 
non-spectator contributions do not play an important role in solving the
missing charm and $\Lambda_b$ lifetime problems.

\noindent {\bf Acknowledgments:} 

This work is partially supported by the NNSF, NSC and ARC.
One of us (Li) would like to thank Dr. Y.L. Wu for discussions.

\begin{table}[t]\caption{The annihilation contributions to the 
branching ratios of $B$ mesons
 for a) $B_1=B_2 = 1$, $\varepsilon_1 = \varepsilon_2 =0$,
and 
b) $B_1 = B_2=1$, $\varepsilon_1 = -0.15$, $\varepsilon_2 = 0$, 
with $m_b = 4.8$ GeV, $f_B= 0.2$ GeV,
 $|V_{ub}|/|V_{cb}| = 0.08$, $|V_{cb}| = |V_{ts}| = 0.038$,
and $|V_{td}| = |V_{ub}|$.\label{tab:exp}}
\vspace{0.4cm}
\begin{center}
\begin{tabular}{|l|l|l|l|l|l|}
\hline 
$\Delta S = 0$&&&&&\\
\hline
$B^-$&a&b&$\bar B^0$&a&b\\
\hline
Tree&&&&&\\
\hline
UN&0&$1.1\times 10^{-5}$&&0&$1.4\times 10^{-4}$\\
PI&$-2.8\times 10^{-4}$&$-7.6\times 10^{-4}$&&0&0\\
\hline
Penguin&&&&&\\
\hline
UN&$1.2\times 10^{-5}$&$1.2\times 10^{-5}$&&0&$1.14\times 10^{-6}$\\
PI&$-0.7\times 10^{-6}$&$-1.4\times 10^{-6}$&&$-0.68\times 10^{-6}$&
$-1.36\times 10^{-6}$\\
\hline
$\Delta S = -1$&&&&&\\
\hline
$B^-$&a&b&$\bar B^0$&a&b\\
\hline
Tree&&&&&\\
\hline
UN&0&$5.7\times 10^{-7}$&&0&0\\
PI&$-1.4\times 10^{-5}$&$-3.9\times 10^{-5}$&&0&0\\
\hline
Penguin&&&&&\\
\hline
UN&$1.9\times 10^{-3}$&$1.9\times 10^{-3}$&&0&0\\
PI&$-1.1\times 10^{-4}$&$-2.2\times 10^{-4}$&&$-1.1\times 10^{-4}$&
$-2.1\times 10^{-4}$\\
\hline
\end{tabular} \end{center} \end{table}

\begin{figure}[htb]
\centerline{ \DESepsf(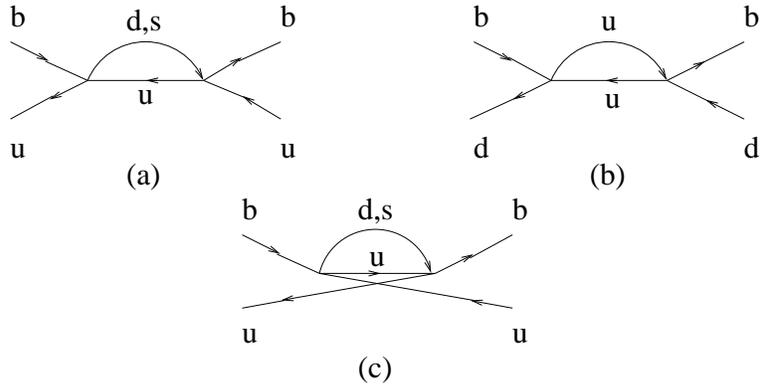 width 10 cm) }
\smallskip
\caption {Diagrams for tree non-spectator contributions to charmless 
$B$ meson decays.
}
\label{tree}
\end{figure}

\begin{figure}[htb]
\centerline{ \DESepsf(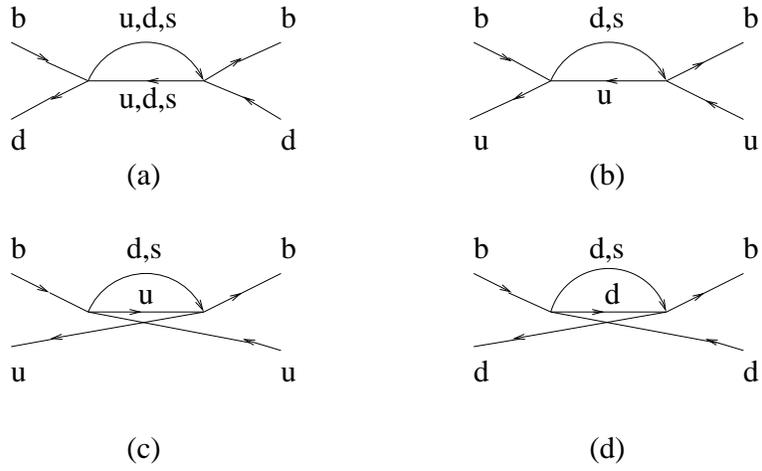 width 10 cm) }
\smallskip
\caption {Diagrams for penguin non-spectator contributions to charmless 
$B$ meson decays.
}
\label{penguin}
\end{figure}
\end{document}